\documentclass[aps,prd,onecolumn,showpacs,groupedaddress,nofootinbib]{revtex4-2}
\usepackage{graphicx}% Include figure files
\usepackage{dcolumn}% Align table columns on the decimal point
\usepackage{bm}% bold math
\usepackage{color}
\usepackage{longtable}
\usepackage{supertabular}
\usepackage[T1]{fontenc}
\usepackage{epstopdf}
\usepackage{amsmath}
\usepackage{amssymb}
\usepackage{setspace}
\usepackage{multirow}
\usepackage{booktabs}
\usepackage[section]{placeins}
\usepackage{mathrsfs}
\usepackage{hyperref}

\makeatletter

\newcommand{\Rmnum}[1]{\expandafter\@slowromancap\romannumeral #1@} 
\newcommand{\bq}{\begin{equation}}
\newcommand{\eq}{\end{equation}}
\newcommand{\bqn}{\begin{eqnarray}}
\newcommand{\eqn}{\end{eqnarray}}
\newcommand{\nb}{\nonumber}

\makeatother

\begin{document}

\title{Utilizing maximum likelihood estimator for flow analysis}

\author{Chong Ye$^{1,2}$}
\author{Wei-Liang Qian$^{3,2,1}$}\email[E-mail: ]{wlqian@usp.br (corresponding author)}
\author{Rui-Hong Yue$^{1}$}
\author{Yogiro Hama$^{4}$}
\author{Takeshi Kodama$^{5,6}$}

\affiliation{$^{1}$ Center for Gravitation and Cosmology, College of Physical Science and Technology, Yangzhou University, Yangzhou 225009, China}
\affiliation{$^{2}$ Faculdade de Engenharia de Guaratinguet\'a, Universidade Estadual Paulista, 12516-410, Guaratinguet\'a, SP, Brazil}
\affiliation{$^{3}$ Escola de Engenharia de Lorena, Universidade de S\~ao Paulo, 12602-810, Lorena, SP, Brazil}
\affiliation{$^{4}$ Instituto de F\'isica, Universidade de S\~ao Paulo, C.P. 66318, 05315-970, S\~ao Paulo-SP, Brazil}
\affiliation{$^{5}$ Instituto de F\'isica, Universidade Federal do Rio de Janeiro, C.P. 68528, 21945-970, Rio de Janeiro-RJ , Brazil}
\affiliation{$^{6}$ Instituto de F\'isica, Universidade Federal Fluminense, 24210-346, Niter\'oi-RJ, Brazil}

\begin{abstract}
We explore the possibility of evaluating flow harmonics by employing the maximum likelihood estimator (MLE).
For a given finite multiplicity, the MLE simultaneously furnishes estimations for all the parameters of the underlying distribution function while efficiently suppressing the variance of measures.
Also, the method provides a means to assess a specific class of mixed harmonics, which is not straightforwardly feasible by the approaches primarily based on particle correlations.
The results are analyzed using the Wald, likelihood ratio, and score tests of hypotheses.
Besides, the resultant flow harmonics obtained using MLE are compared with those derived using particle correlations and event plane methods.
The dependencies of extracted flow harmonics on the multiplicity of individual events and the total number of events are analyzed.
It is shown that the proposed approach works efficiently to deal with the deficiency in detector acceptability.
Moreover, we elaborate on a fictitious scenario where the event plane is not a well-defined quantity in the distribution function.
For the latter case, the MLE is shown to largely perform better than the two-particle correlation estimator. 
In this regard, one concludes that the MLE furnishes a meaningful alternative to the existing approaches for flow analysis.

\end{abstract}

%\date{\today}
\date{Oct. 12th, 2022}

\maketitle
\newpage

\section{Introduction}\label{section1}

Relativistic hydrodynamics is one of the feasible theoretical frameworks to describe the temporal evolution of the strongly coupled quark-gluon plasmas created in the relativistic heavy-ion collisions~\cite{hydro-review-04, hydro-review-05, hydro-review-06, hydro-review-07, hydro-review-08, hydro-review-09, hydro-review-10}.
As a macroscopic approach, the plasma is modeled as a continuum, essential in understanding the underlying physics leading to the empirical observables. 
The main goal of relativistic hydrodynamics is to describe the observed particle spectra at intermediate and low transverse momentum, notably their collective properties, such as the flow harmonics and correlations~\cite{Ollitrault:1992bk, Voloshin:1994mz, Ollitrault:1997vz,  Borghini:2000sa, Takahashi:2009na, Andrade:2010xy, Luzum:2010sp}. 
% Generally, a hydrodynamic approach utilizes a specific microscopic model, which furnishes the initial conditions as its input. 
%The initial conditions are primarily expressed in terms of the density distribution, while specific microscopic details of the initial state are largely washed out during the pre-equilibrium stage, for which one typically assumes that the thermalization is attained. 
%Afterward, the fluid expands according to the ideal or viscous relativistic hydrodynamic equations until the hadronization occurs.
From the experimental perspective, the measured azimuthal distributions played a crucial role in establishing the picture of a ``perfect'' liquid, first served up at RHIC~\cite{STAR:2000ekf}. 
In turn, the analysis of the azimuthal anisotropy produced by nuclear collisions has become one of the most prominent observables to extract crucial information on the properties of the underlying physical system~\cite{BRAHMS:2004adc, PHOBOS:2004zne, STAR:2005gfr, ATLAS:2012at, CMS:2012zex, CMS:2012tqw}.

The anisotropic distribution of particles in the momentum space is expressed in the flow harmonics $v_n$. 
To be specific, we have the following one-particle distribution function~\cite{Voloshin:1994mz, Voloshin:2008dg}
\begin{eqnarray}
f_1(\phi)=\frac{1}{2\pi}\left[1+\sum_{n=1}2v_{n}\cos{n(\phi-\Psi_n)}\right] ,
\label{oneParDis}
\end{eqnarray}
where $\phi$ represents the azimuth angle of an emitted particle, and the reference orientation $\Psi_n$ for a given order $n$ is referred to as the event plane. 
Compared to others, the elliptic flow $v_2$ and triangular flow $v_3$ account for many significant features of the observed system.
In particular, $v_2$ is due mainly to the geometric almond shape of the initial system~\cite{Ollitrault:1992bk}, and $v_3$ is attributed to the event-by-event fluctuations of the initial conditions~\cite{Alver:2010gr}.
In the literature, much effort has been devoted to investigating the relationship between the initial geometric anisotropy and the final-state flow harmonics, and in particular, the deviation of the underlying relation from a linearized one~\cite{Teaney:2012ke, Niemi:2012aj, Qian:2013nba, Yan:2014nsa, Fu:2015wba, Yan:2015jma, Wen:2020amn}, eccentricity and flow fluctuations~\cite{Hama:2007dq, Bhalerao:2011yg, Heinz:2013bua, Gronqvist:2016hym}, and correlations~\cite{Bhalerao:2013ina, Denicol:2014ywa}.%\blue{(xxx Bhalerao here is fit or not xxx)}. 

In practice, many different approaches have been developed to numerically determine the value of flow harmonics $v_{n}$ from the experimental data.
The conventional event plane method~\cite{Voloshin:1994mz, Poskanzer:1998yz} aims to estimate the event planes $\Psi_n$ in Eq.~\eqref{oneParDis} and then evaluate the flow harmonics.
The main feature of the approach is closely related to the fact that the reaction plane~\cite{Alver:2010gr} cannot be directly measured experimentally.
Many other methods are primarily based on the particle correlations and the notions of the Q-vectors and cumulants~\cite{Danielewicz:1985hn, Borghini:2000sa, Bilandzic:2010jr,  Jia:2017hbm}.%\blue{(1985,2000 is about Q-vector, 2010 about cumulants)}.
One advantage of using particle correlation is that the event planes in Eq.~\eqref{oneParDis} are canceled out in the formalism.
Moreover, the cumulant can be expressed concisely in terms of the generating function~\cite{Borghini:2000sa, Borghini:2001vi}.
It is worth noting that the notation of multi-particle cumulants may evolve harmonics of a different order, as discussed in several relevant studies. 
This class of approaches include the particle cumulant~\cite{Borghini:2000sa, Borghini:2001vi, Bilandzic:2010jr}, Lee-Yang zeros~\cite{Bhalerao:2003xf, Bhalerao:2003yq, FOPI:2005ukb}, and symmetric cumulants~\cite{Bilandzic:2013kga}, among others recent generalizations~\cite{ Bhalerao:2013ina, DiFrancesco:2016srj, Mordasini:2019hut}.

There are two essential aspects of flow analysis.
The first one is non-flow, a generic term used to refer to the collective behavior that cannot be attributed to independent emission based on the one-particle distribution function Eq.~\eqref{oneParDis}.
For instance, momentum conservation leads to deviation~\cite{Chajecki:2008yi} in particle correlation compared to the case where the particle spectrum is entirely governed by the flow.
Generally, it is understood that the impact due to non-flow becomes more insignificant as one considers particle correlation involving more hadrons~\cite{Borghini:2001vi, Borghini:2001zr} or introduces a rapidity gap among the particles~\cite{PHENIX:2003qra, Voloshin:2006wi, Voloshin:2006gz}.
The second one is related to the finite multiplicity in individual events.
As a matter of fact, even though the hadrons are emanated independently according to the one-particle distribution function, the second factor leads to a certain degree of statistical uncertainty.
Such uncertainty is distinct from those caused by the fluctuations in the initial conditions, which typically possess some physical origin regarding the underlying microscopic model.
To be precise, from a hydrodynamic perspective, the latter is manifestly by the fluctuations in the geometry of the initial energy distribution on an event-by-event basis.
In particular, the statistical error of the particle correlation method might even be larger than that of the event plane method~\cite{Voloshin:2008dg}. 
Besides, for events with lower multiplicity, the multi-particle correlation is likely constrained by more significant statistical uncertainty.

To clarify the above statement, let us consider the following simple example.
By assuming that the flow is constituted by independent particle emission according to 
\begin{eqnarray}
    f(\phi_1, \phi_2)=f_1(\phi_1) f_1(\phi_2), \label{findf1}
\end{eqnarray}
where $f_1$ is defined by Eq.~\eqref{oneParDis}, one has
\begin{eqnarray}
    \langle 2 \rangle_{2,-2} \equiv \langle e^{i2(\phi_{1}-\phi_{2})}\rangle 
    = \langle\cos{2(\phi_{1}-\phi_{2})}\rangle + i \langle\sin{2(\phi_{1}-\phi_{2})}\rangle = v_2^2, \label{v22inCumu}
\end{eqnarray}
where $\langle\cdots\rangle\equiv \int d\phi_1d\phi_2 f(\phi_1,\phi_2)\cdots$ is evaluated by averaging out all the particle pairs of a given event, in accordance with the probability given by the one-particle distribution function, at the limit of infinity multiplicity. 
One notes that the imaginary part of Eq.~\eqref{v22inCumu} vanishes by employing the orthogonal relationship of sinusoidal functions.
On the other hand, for an event of a finite number of multiplicity $M$, it is natural to estimate $v_2^2$ by a discrete version of Eq.~\eqref{v22inCumu}, namely,
\begin{eqnarray}
    \widehat{v_2^2}\equiv \frac{1}{M(M-1)}\sum_{\substack{i \ne j}}\cos{2(\phi_{i}-\phi_{j})} ,
\label{Estimatorv2c2}
\end{eqnarray}
where the summation enumerates all distinct $M(M-1)$ ordered pairs. 
In this regard, one can view Eq.~\eqref{Estimatorv2c2} as a {\it statistical estimator}~\cite{book-statistics-Rice} $\hat{\theta}$ of a physical quantity $\theta = v_2^2$ for an individual event.
Given the underlying distribution, as a statistical estimator, its {\it quality} is usually measured in terms of its expected value $\mu$ and variance $\sigma^2$.
In the case of Eq.~\eqref{Estimatorv2c2}, one has
\begin{eqnarray}
    \mu &\equiv& \mathrm{E}\left[\widehat{v_2^2}\right]= v_2^2 ,\label{muv22}\\ 
    \sigma^2 &\equiv& \mathrm{Var}\left[\widehat{v_2^2}\right] 
    = \frac{1+v_{2}^2}{M(M-1)}+2\frac{M-2}{M(M-1)}v_{2}^{2}(1+v_{4}) + \frac{(M-2)(M-3)}{M(M-1)}v_{2}^{4}-v_{2}^{4}. \label{sigma2v22}
\end{eqnarray}
Here, Eq.~\eqref{muv22} indicates that the estimator Eq.~\eqref{Estimatorv2c2} is unbiased.
On the other hand, Eq.~\eqref{sigma2v22} clearly shows that the statistical uncertainty does not vanish as long as the multiplicity $M$ is finite.
Different from Eq.~\eqref{Estimatorv2c2}, Eq.~\eqref{v22inCumu} evaluates the expectation at the limit of infinite multiplicity.
On the other hand, any given flow evaluation scheme must be applied to realistic events with finite multiplicity and, therefore, can always be viewed as a specific choice of statistical estimator.
As a result, it is inevitably subject to some statistical uncertainty.

The present study is mainly motivated to explore the above aspect in flow measurements.
We examine the possibility of utilizing the maximum likelihood estimator (MLE) as a flow estimator.
The MLE is a well-known approach to estimating the parameters of a hypothetical probability distribution when observed data is given.
Its goal is achieved by maximizing the associated likelihood function in the parameter space.
In other words, the estimated parameters, for our case, the flow harmonics, ensure that the observed data is most probable.
The method is widely used in statistical inference due to its intuitive and flexible nature.
In particular, as an asymptotically normal estimator, its efficiency is guaranteed in the sense that it is more accurate than any other estimator at the limit of a significant sample size.  
In terms of convergence, the method is either unbiased or asymptotically unbiased. 
Apparently, the context of relativistic heavy ion collisions meets most of the characteristics of MLE. 
In particular, the measurements performed at RHIC and LHC have accumulated significant events for different collision systems at different centralities.

The remainder of the present paper is organized as follows. 
In the next section, after briefly reviewing the conventional particle correlation method, we discuss the mathematical framework of the MLE and its application to flow analysis in relativistic heavy-ion collisions. 
In Sec.~\ref{section3}, we carry out numerical studies to illustrate the use of the MLE method based on the simulated events using Monte Carlo.
The results are then compared with the particle correlation method.
The method's efficiency and dependence on the multiplicity and number of events are analyzed. 
In Sec.~\ref{section4}, we apply the method to the scenario where the detector inefficiencies play a part.
The results are again compared with those obtained using the multi-particle correlation method in terms of the Q-vectors.
Last but not least, in Sec.~\ref{section5}, we elaborate on a fictitious scenario where the event plane is not a well-defined quantity in the distribution function.
The last section is devoted to further discussions and concluding remarks. 

\section{Statistical estimators for flow harmonics}\label{section2}

\subsection{Measurement of flow harmonics using particle correlations}

As mentioned above, the most prominent approaches to extract the flow harmonics are based on particle correlations~\cite{Poskanzer:1998yz}.
The cornerstone of such an approach is based on the following relation regarding $k$-particle correlation~\cite{Bhalerao:2011yg} %\blue{(high order correlations)}
\begin{eqnarray}
    \langle k\rangle_{n_1,\cdots,n_k}\equiv \langle e^{i(n_1\phi_1 + \cdots + n_k\phi_k)} \rangle 
    = v_{n_1}\cdots v_{n_k} e^{i\left(n_1\Psi_{n_1}+\cdots+n_k\Psi_{n_k}\right)} ,  \label{nkCorr}
\end{eqnarray}
where $\langle\cdots\rangle$ is an average over distinct tuples of particles, again assuming independent particle emission according to Eq.~\eqref{oneParDis} particles of a given event, at the limit of infinity multiplicity.

To concentrate on $v_n$, one usually chooses a specific set of $(n_1, \cdots, n_k)$ so that $\sum_{j=1}^k n_j=0$~\cite{Bhalerao:2011yg}, and therefore, all the coefficients involving the event planes cancel out in the exponential.
Otherwise, the event plane correlator will become a part of the formalism~\cite{Bhalerao:2013ina}.
For instance, in the case of two-particle correlation $k=2$, one may consider $n_1 = -n_2 = n$.
Therefore, we have
\begin{eqnarray}
        \langle 2 \rangle_{n,-n} \equiv \left \langle e^{in(\phi_{1}-\phi_{2} )}\right \rangle 
        =  \langle\cos{n(\phi_{1}-\phi_{2})} \rangle = v_n^2
\label{eq2}
\end{eqnarray}
For an event with finite multiplicity $M$, where the azimuth angles of the measured particles read $\phi_{1}, \phi_{2}, \cdots, \phi_{M}$, it is intuitive to adopt the following estimator~\cite{Bilandzic:2013kga}
\begin{eqnarray}
        \widehat{v_n^2} =\frac{1}{M(M-1)}\sum_{i\ne j}\cos n(\phi_{i}-\phi_{j}).
\label{eqEst2}
\end{eqnarray}
Subsequently, the first two moments of the estimator are found to be~\cite{Bilandzic:2013kga}
\begin{eqnarray}
      \mathrm{E}\left[\widehat{v_n^2}\right] &=& v_{n}^{2},\label{eqE3}\\
      \mathrm{Var}\left[\widehat{v_n^2}\right] &=& \frac{1+v_{2n}^2}{M(M-1)}+2\frac{M-2}{M(M-1)}v_{n}^{2}(1+v_{2n})     +\frac{(M-2)(M-3)}{M(M-1)}v_{n}^{4}-v_{n}^{4} ,\label{eqVar3}
\end{eqnarray}
which is an unbiased estimator that has a finite variance that decreases with increasing multiplicity.
This result readily falls back to Eqs.~\eqref{muv22} and~\eqref{sigma2v22} given above.

To proceed, one may either proceed to evaluate cumulant using the formalism of generating functions, as first proposed by Borghini {\it et al.}~\cite{Borghini:2000sa, Borghini:2003ur, Borghini:2007ku}, or evaluate the multi-particle correlation straightforwardly~\cite{Bilandzic:2010jr}. %\blue{use generating functions to calculate multi-particle correlations)}

The above formalism can also be generalized to include weighted average~\cite{Bilandzic:2013kga}, where a specific weight $w_k$ is associated with $k$th particle in the correlation.
In particular, we have
\begin{eqnarray}
     \langle k \rangle_{n_{1},n_{2}\cdots n_{k}} 
     \equiv \langle e^{i(n_{1}\phi_{1}+n_{2}\phi_{2}+\cdots+n_{k}\phi_k)} \rangle ,
\label{MultiKeq4}
\end{eqnarray}
which is intuitively given by
\begin{eqnarray}
\frac{\sum\limits_{k-\mathrm{tuples}} w_{1}w_{2}\cdots w_{k} e^{i(n_{1}\phi_1+n_{2}\phi_2+\cdots+n_{k}\phi_k)}}{\sum\limits_{k-\mathrm{tuples}} w_{1}w_{2}\cdots w_{k}}
    \equiv \frac{\mathrm{N}_{\langle k \rangle_{n_{1},n_{2}\cdots n_{k}}}}{\mathrm{D}_{\langle k \rangle_{n_{1},n_{2}\cdots n_{k}}}} ,
\label{MultiK_dis}
\end{eqnarray}
at finite multiplicity, where the summation is carried out for all distinct tuples, from which any auto-correlation should be removed.

In practice, the numerator and denominator of Eq.~\eqref{MultiK_dis} can be expressed by employing the Q-vectors~\cite{Danielewicz:1985hn}, defined as
\begin{eqnarray}
    Q_{n,p}\equiv \sum_{j=1}^{M}w_{j}^{p}e^{in\phi_{j}} ,
\label{qVec}
\end{eqnarray}
where $p$ is an exponent that can be chosen conveniently to simply the resultant expressions.
As an example, for $k=2$, one has
\begin{eqnarray}
     \mathrm{N}_{\langle 2\rangle _{n_1, n_2}} &=& Q_{n_1,1}Q_{n_2,1}-Q_{n_1+n_2,2} , \label{eqN8}\\
     \mathrm{D}_{\langle 2\rangle _{n_1, n_2}} &=& Q_{0,1}^{2}-Q_{0,2} ,\label{eqD8}
\label{eqQvector2p}
\end{eqnarray}
which falls back to (the real part of) the r.h.s. of Eq.~\eqref{eqEst2} for $w_j=1$ and $n_1=-n_2=n$.

Again, we note that Eqs.~\eqref{MultiK_dis} and~\eqref{eqN8} are essentially estimators for a given event with finite multiplicity obtained experimentally, which, in turn, are subject to statistical uncertainties.
To make the above generic statements concrete, we consider two simple examples.
For an event with multiplicity $M$, let $k=2$, $w_1= w_2 =1$ and $n_1=-n_2=n$, we have
\begin{eqnarray}
      \mathrm{E}\left[\mathrm{N}_{\langle 2\rangle _{n_1, n_2}}\right] 
      &=& {M(M-1)v_{n}^{2}} ,\\
      \mathrm{Var}\left[\mathrm{N}_{\langle 2\rangle _{n_1, n_2}}\right] 
      &=& M(M-1)\left\{(1+v_{2n}^2)+2(M-2)v_{n}^{2}(1+v_{2n})-2(2M-3)v_{n}^{4}\right\} .\label{eqVar3a}
\end{eqnarray}
As a second example, consider $w_1 = w_2 = w_3 =1$, $n_1 =2$, $n_2=3$, $n_3=-5$, and let us assume $\Psi_2=\Psi_3=\Psi_5=\Psi$ for simplicity, one finds
\begin{eqnarray}
     \mathrm{E}\left[\mathrm{N}_{\langle 3\rangle _{2, 3, -5}}\right] 
      &=& {M(M-1)(M-2)v_2v_3v_5} ,\label{p3N}\\
      \mathrm{Var}\left[\mathrm{N}_{\langle 3\rangle _{2, 3, -5}}\right] 
      &=& M(M-1)(M-2)\left\{\left[v_4v_6v_{10}+v_2^2v_4+v_3^2v_6+v_1^2v_{10}+2v_2v_3v_5\right]\right.\nb\\
      &+& 2(M-3)\left[\left(v_2^2v_5^2+v_2v_3v_5v_6\right)+\left(v_2v_3v_4v_5+v_3^2v_5^2\right)+\left(v_2^2v_3^2+v_2v_3v_5v_{10}\right)\right] \nb\\
      &+& (M-3)\left[\left(v_3^2v_4v_{10}+v_3^4\right)+\left(v_2^2v_6v_{10}+v_2^4\right)+\left(v_4v_5^2v_6+v_5^4\right)\right] \nb\\
      &+& (M-3)(M-4)\left[2v_2v_3v_5^3+2v_2v_3^3v_5+2v_2^3v_3v_5+v_3^2v_4v_5^2+v_2^2v_5^2v_6+v_2^2v_3^2v_{10}\right] \nb\\
      &+& \left.(M-3)(M-4)(M-5)v_2^2v_3^2v_5^2- M(M-1)(M-2)v_2^2v_3^2v_5^2\right\} .\label{eqVar3b}
\end{eqnarray}
Although a bit tedious, evaluating the last expression is straightforward by considering different combinations when one, two, or three particles from the two ordered triple (consisting of distinct particles, respectively) coincide.
Eqs.~\eqref{eqVar3a} and~\eqref{eqVar3b} illustrate the fact that these estimators are subject to finite uncertainties and do not vanish, owing to finite statistics.
The event planes do not coincide, and therefore, the r.h.s. of the three-particle correlation Eq.~\eqref{p3N} actually furnishes the event plane correlation~\cite{Bhalerao:2011yg}.

In this regard, one might speculate that MLE could serve as an alternative flow estimator, and such a possibility will be explored in the remainder of the paper.
The definition and elementary properties of the MLE are elaborated on in the following subsection. 
Subsequently, the approach is implemented numerically.
The Monte Carlo simulations are performed, and the results will be compared to the conventional techniques, such as the particle correlation and event-plane methods.

\subsection{MLE and its application as a flow estimator}

In this subsection, we turn our attention to the MLE.
For a given set of observations $y\equiv (y_1, y_2, \cdots, y_M)$, we assume that they are sampled from a joint probability distribution governed by several unknown parameters $\theta \equiv (\theta_1, \theta_2, \cdots, \theta_m)$.
As mentioned in the introduction, one considers the following likelihood function $\mathcal{L}$ at the observed data 
\begin{eqnarray}
     \mathcal{L}(\theta) \equiv \mathcal{L}(\theta; y) = f(y; \theta).\label{defLn}
\end{eqnarray}
which is the joint probability density for the given observation evaluated at the parameters $\theta$.
The goal of MLE is to determine the parameters for which the observed data attains the highest joint probability, namely
\begin{eqnarray}
     \hat{\theta}_{\mathrm{MLE}}=\arg\max\limits_{\theta\in\Theta}\mathcal{L}(\theta) , \label{defMLE}
\end{eqnarray}
where $\Theta$ is the domain of the parameters.
In particular, for independent and identically distributed (i.i.d.) random variables, $f(y; \theta)$ is given by 
\begin{eqnarray}
     f(y; \theta)=\prod_{j=1}^M f^\mathrm{uni}(y_j; \theta) . \label{iidfn1}
\end{eqnarray}

Indeed, the above scheme can be readily applied in the context of collective flow in heavy-ion collisions.
Considering an event consisting of $M$ particles, the likelihood function reads 
\begin{eqnarray}
\mathcal{L}(\theta; \phi_{1}, \cdots, \phi_M) = f(\phi_{1},\cdots, \phi_M; \theta)=\prod_{j=1}^{M}f_1(\phi_j; \theta) ,
\label{eqlikelihood}
\end{eqnarray}
where the likelihood function $\mathcal{L}$ is governed by the one-particle distribution function Eq.~\eqref{findf1}.
The last equality is based on the assumption that the particles' azimuthal angles are i.i.d. variables, which gives Eq.~\eqref{findf1} in the case of two particles.
The parameters of the distribution, $\theta = (v_1, \Psi_1, v_2, \Psi_2, \cdots)$, are the flow harmonics and event planes.

In practice, one often chooses the objective function to be the log-likelihood function ${\ell}$
\begin{eqnarray}
{\ell}(\theta; \phi_{1}, \cdots, \phi_M) = \log\mathcal{L}(\theta; \phi_{1}, \cdots, \phi_M)
  = \sum_{j=1}^{M}\log f_1(\phi_j; \theta) .
\label{eqlogl}
\end{eqnarray}
Numerical calculations indicate that Eq.~\eqref{eqlogl} is more favorable than Eq.~\eqref{eqlikelihood}, although as the multiplicity $M$ becomes more significant, the appropriate implementation should be adopted to avoid increasing truncation error.

The maximum of $\ell$ occurs at the same value of $\theta$ as does the maximum of $\mathcal{L}$.
For $\ell$ that is differentiable in its domain $\Theta$, the necessary conditions for the occurrence of a maximum are
\begin{eqnarray}
\frac{\partial{\ell}}{\partial\theta_1}=\cdots=\frac{\partial{\ell}}{\partial\theta_m}=0 .
\label{condMLE}
\end{eqnarray}

In this work, we will primarily focus on the dominant harmonic coefficients such as $v_2$ and $v_3$ even though the associated event planes are obtained simultaneously from Eq.~\eqref{condMLE}.

As discussed in the introduction, MLE has asymptotical normality, which attains the Cramér-Rao lower bound when the sample size increases.
In other words, no consistent estimator has a lower asymptotic mean squared error than the MLE.
In the context of relativistic heavy-ion collisions, all the events of a given multiplicity asymptotically form a (multivariant) normal distribution
\begin{eqnarray}
\hat{\theta}_{\mathrm{MLE}} \sim N\left(\theta_0, (I_M(\theta_0))^{-1}\right) ,
\label{assNormMLE}
\end{eqnarray}
where $\theta_0$ represents the ``true'' value, and $I_M(\theta)$ is the Fisher information matrix, defined as
\begin{eqnarray}
I_M(\theta) \equiv E_\theta\left[-\frac{d^2}{d\theta^2}{\ell}(\theta;\phi_1,\cdots,\phi_M)\right] ,
\label{defIM}
\end{eqnarray}
where the expectation $E_\theta$ is taken with respect to the distribution of $f(\phi_1,\cdots,\phi_M;\theta)$.
For i.i.d. data, the Fisher information possesses the form
\begin{eqnarray}
I_M(\theta) = M I_1(\theta),
\label{defI1}
\end{eqnarray}
where $I_1$ is the Fisher information matrix for a single observation.
As a result, the standard deviation of MLE is expected to be roughly proportional to $\frac{1}{\sqrt{M}}$.
These properties can be further quantified using the Wald, likelihood ratio, and score tests, which will be performed in the numerical simulations. 

\section{Monte Carlo simulations}\label{section3}

\subsection{Numerical results and comparison with other methods}

Based on the discussions in the preceding sections, we employ the Monte Carlo generator to simulate events and estimate the flow coefficients using the MLE method.
The quality of the estimation is evaluated by the standard tests of hypotheses.
Also, the results are compared with those obtained using the other approaches, namely, the particle correlation and event plane methods.

One primarily considers events consisting of independent particle emissions according to the one-particle distribution function Eq.~\eqref{oneParDis}, which is further truncated at the fourth order, and the directed flow is ignored 
\begin{eqnarray}
f_1(\phi)=\frac{1}{2\pi }\left[1+2v_2\cos{2(\phi-\Psi_2)}+2v_3\cos{3(\phi-\Psi_3)}+2v_4\cos{4(\phi-\Psi_4)}\right] ,
\label{eqfphiv234}
\end{eqnarray}
where the event planes $\Psi_n$ are randomized between distinct events, and the harmonics are taken as
\begin{eqnarray}
v_2&=&0.2, \nonumber\\
v_3&=&0.08, \nonumber\\
v_4&=&0.25.
\label{eqvnvalue}
\end{eqnarray}

We use the above parameters to generate a specific number of events with a given multiplicity.
The MLE method is then employed to extract the flow harmonics. 
To validate the MLE approach, we aim to see whether the estimation of $v_2$ will be undermined by some other parameter that plays a more significant role in the underlying probability distribution.
In this regard, we have particularly chosen a large and unphysical value for $v_4$.
We have verified that the results below are not sensitive to this specific choice.
Besides, for simplicity, we do not consider the flow's momentum or rapidity dependence; therefore, only the particle's azimuthal angles are involved in the calculations. 

As a first example, we consider one single event with a multiplicity $M=1000$.
In Tab.~\ref{tabvn1e}, we show the estimated parameters.
The corresponding results using the Wald, likelihood ratio, and score tests are shown in Tab.~\ref{Wtests}.

\begin{table}
\caption{The estimated flow harmonics $v_n$ and event planes $\Psi_n$ using MLE from a single event.}\label{tabvn1e}
\begin{tabular}{c ccc ccc}
        \hline\hline
            $\theta$          & $v_2$         & $v_3$           & $v_4$  & $\Psi_2$ &  $\Psi_3$ & $\Psi_4$\\
          \hline
          true values  &  0.2       & 0.08         & 0.25  &  2.25      & 0.0323        & 0.925 \\
          \hline
         MLE          &  0.228   &  0.0782   & 0.253 &  2.36   &  0.0461      & 0.966\\
         \hline\hline
\end{tabular}
\end{table}
 
\begin{table}
\caption{The results of the Wald, likelihood ratio, and score tests of hypotheses.}\label{Wtests}
\begin{tabular}{ cccccc}
        \hline\hline
                        Wald           & likelihood ratio     & score\\
        \hline
             4.981       & 2.061         & 6.394  \\
         \hline\hline
\end{tabular}
\end{table}

The quality of the estimation is further evaluated by the statistical tests of hypotheses~\cite{neyman1952lectures}, which are presented in Tab.~\ref{Wtests}.
In particular, we perform three types of hypothesis testing: Wald, likelihood ratio, and score tests.
Roughly speaking, all these tests assess constraints on statistical parameters by quantifying the deviation between the unrestricted estimate and its hypothesized value.
Analysis such as confidence intervals for the estimator can be facilitated by the fact that all three tests asymptotically approach the $\chi^2$ distribution, a sum of the squares of normal distributions under the null hypothesis.
Intuitively, in terms of the log-likelihood function, one measures the precision of the estimation in terms of the horizontal and vertical distances between the estimator $\hat{\theta}$ and the true value ${\theta}_0$.
Specifically, the Wald test access how far away the estimator locates in the horizontal direction, and the likelihood ratio test gives that in the vertical direction, namely,
\begin{eqnarray}
t_\mathrm{W} =\left(\hat{\theta}-\theta_0\right)^T \hat{V}_M^{-1}\left(\hat{\theta}-\theta_0\right) ,
\label{tWald}
\end{eqnarray}
and
\begin{eqnarray}
t_\mathrm{LR} =2\left[\ell(\hat{\theta})-\ell({\theta}_0)\right] .
\label{tLR}
\end{eqnarray}
where the covariant matrix $\hat{V}_M $ is
\begin{eqnarray}
\hat{V}_M =\left[I_M(\theta_0)\right]^{-1}  ,
\label{hatvn}
\end{eqnarray}
where $I_M$ is given by Eq.~\eqref{defIM}, the Fisher information under the null hypothesis and $\ell$ is the log-likelihood function, and the factor ``2'' ensures that it converges asymptotically to a $\chi^2$-distribution.

The score test measures the slope at the true value
\begin{eqnarray}
t_\mathrm{S} =S(\theta_0)^T\hat{V}_M S(\theta_0) ,
\label{tScore}
\end{eqnarray}
where 
\begin{eqnarray}
S(\theta) =\nabla_\theta \ell(\theta;\phi_1, \cdots,\phi_M) ,
\label{defS}
\end{eqnarray} 
is the gradient of the log-likelihood function,

For the specific event, the Fisher information $I_1(\theta_0)$ is numerically found to be
\begin{equation} 
\left(                
  \begin{array}{cccccc} 
    2.148  & 0.479  & 0.106 & -0.247 & -0.015 & 0.579\\ 
    0.479  & 3.140  & 0.479 & -0.107 & -0.038 & 0.239\\  
    0.106  & 0.479  & 2.819 & -0.208 & -0.065 & 0.084\\ 
    -0.247 & -0.107 & -0.208 & 0.655 & 0.024  & -0.348\\ 
    -0.015 & -0.038 & -0.065 & 0.024 & 0.155  & 0.004\\ 
    0.579  & 0.239  & 0.084 & -0.348 & 0.004  & 2.915\\ 
  \end{array}
\right) ,           
\end{equation}
where the dimension of the matrix is determined by that of $\theta$. 
The MLE results manifestly sit within the 95\% confident interval.

We have observed that, even for a single event, the flow harmonics estimated by the MLE method are reasonable.
In what follows, we compared the obtained flow harmonics with those extracted using particle correlations.
Also, since the method gives a more favorable asymptotic mean squared error, it might be used to achieve a minimized statistical uncertainty when compared to other approaches.
In this regard, we consider a scenario where the elliptic flow fluctuates on an event-by-event basis.
To be specific, one replaces the first line of Eq.~\eqref{eqvnvalue} with the following Gaussian distribution
\begin{eqnarray}
v_2 \sim N(\mu=0.2, \sigma^2=0.01) .
\label{v2sigma}
\end{eqnarray}

\begin{figure}[ht]
\begin{tabular}{cc}
\vspace{-26pt}
\begin{minipage}{250pt}
\centerline{\includegraphics[width=1.2\textwidth,height=1.0\textwidth]{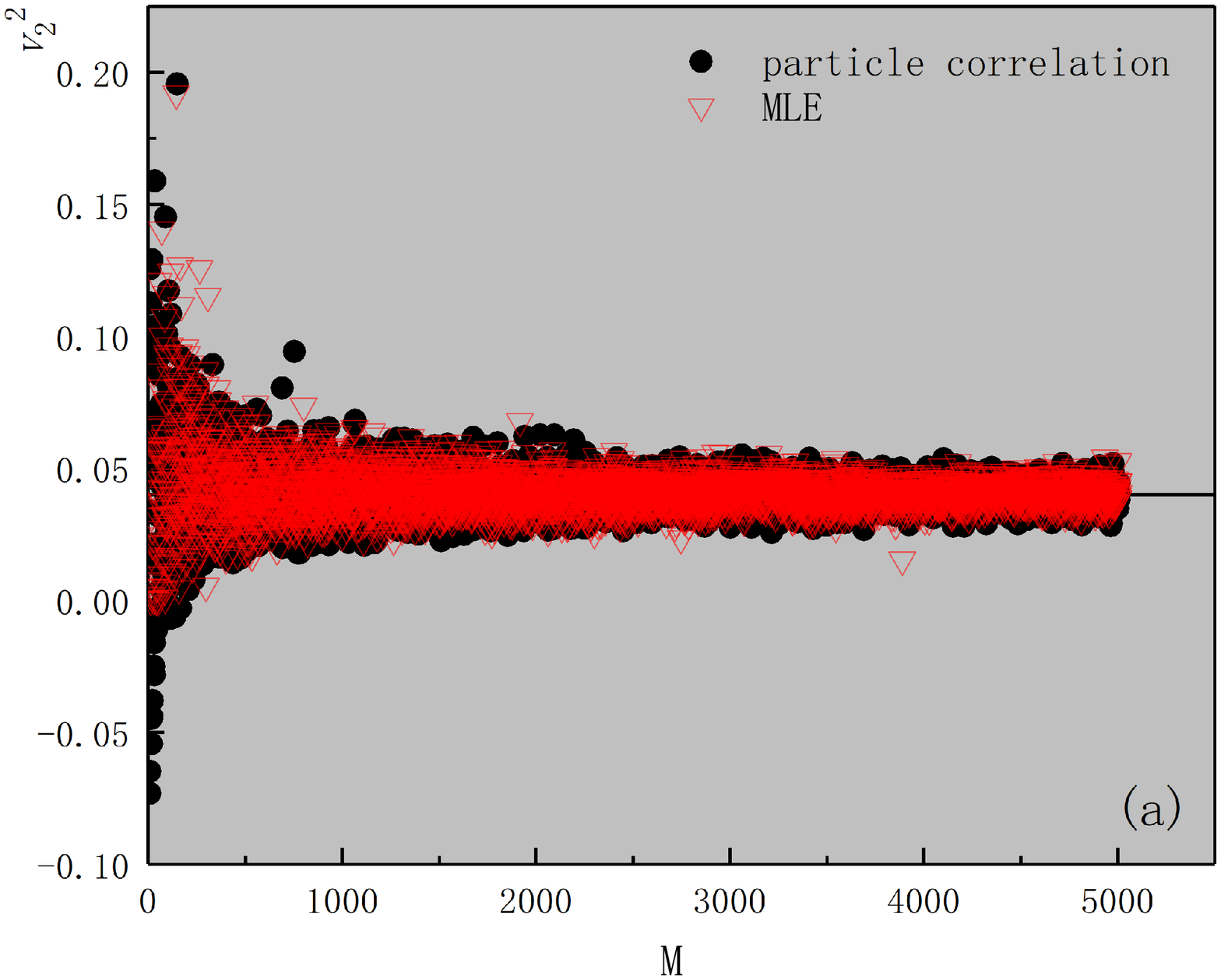}}
\end{minipage}
&
\vspace{22pt}
\begin{minipage}{250pt}
\centerline{\includegraphics[width=1.2\textwidth,height=1.0\textwidth]{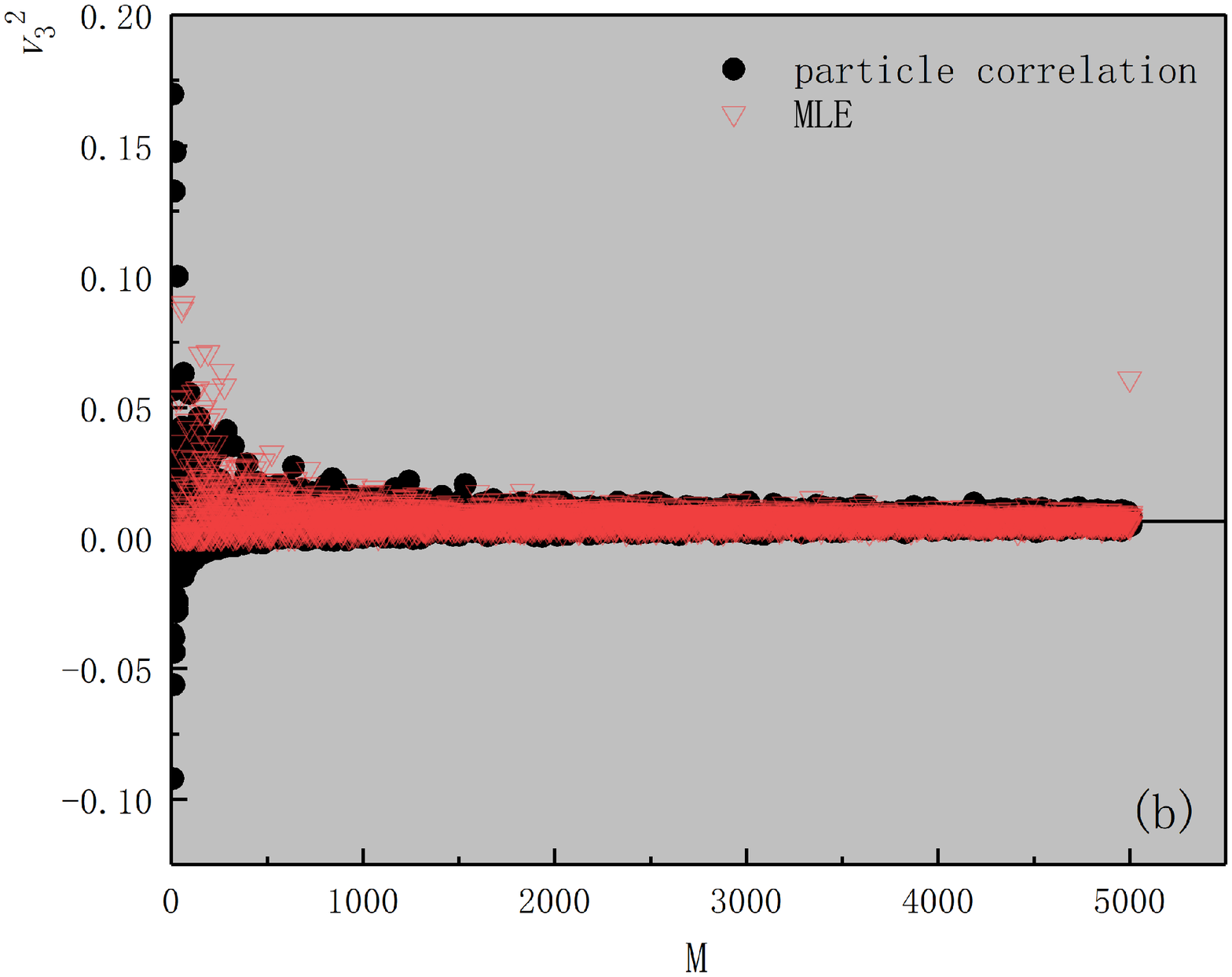}}
\end{minipage}
\end{tabular}
\vspace{12pt}
\renewcommand{\figurename}{Fig.}
\caption{(Color Online) The estimated flow harmonics $v_2^2$ and $v_3^2$ using the MLE and particle correlation methods, as functions of the event multiplicity $M$.
The hollow red triangles represent the results obtained by the MLE, while the filled black circles are those by the particle correlation method.}
\label{figv2v3scatter}
\end{figure}

The numerical results on the comparison between the two-particle correlation and MLE methods are presented in Tabs.~\ref{tabv2vara} and~\ref{tabv2varb}.
First, for each event of a given multiplicity, particles are drawn independently according to Eq.~\eqref{eqfphiv234} using the parameters given by Eq.~\eqref{eqvnvalue} and estimate the $v_2$ for individual events by the MLE and then calculate the average.
In the calculations, for simplicity, we adopt the values estimated by the event-plane method for the remaining flow harmonics and the event planes.
This applies to the results presented in Tabs.~\ref{tabv2vara},~\ref{tabv2varb},~\ref{tabeventinc1000}, and Fig.~\ref{figv2v3scatter} in the current subsection, and also to Figs.~\ref{piecefcorr}-\ref{figv234corr} below in Sec.~\ref{section4}, as well as to Tab.~\ref{tabpsi2varc} in Sec.~\ref{section5}.
For the remaining applications of MLE, the dimension of the parameter space $\theta$ is six-dimensional.

The resulting elliptic flow is obtained by the event average of MLE estimations and shown in Tab.~\ref{tabv2vara}.
The corresponding variance of $v_2$ among different events, a measure of flow fluctuations, is also evaluated.
The results obtained by the MLE method are compared with those using the two-particle correlation formula Eq.~\eqref{eqEst2}. 
Specifically, for the latter, the elliptic flow is estimated by taking the square root of the event average of the r.h.s. of Eq.~\eqref{eqEst2}, in accordance with the practice~\cite{Nadderd:2021fst}.
Also, the variance of the flow is evaluated by that of $\sqrt{\widehat{v_2^2}}$ for individual events, following Refs.~\cite{STAR:2010uoq}. 
It is found that both methods give a reasonable estimation of the flow harmonics.
As the multiplicity $M$ increases, the variance extracted from the MLE method becomes smaller.
It is noted that the values of the mean and variance numerically converge as long as one uses a significant number of events.
The results presented in Tab.~\ref{tabv2vara} are obtained using $N=1000$ events.
We also note that in this case, the origin of the variance of the elliptic flow is entirely statistical because $v_2$ in the one-particle distribution is well-defined.
In this sense, the column of $\mathrm{Var}[v_2^2]$ denoted by ``true value'' is filled with zeros.

For a realistic scenario, on the other hand, it is understood that the measured flow fluctuations also contain the part which is governed by the event-by-event fluctuations of initial geometry.
The latter reflects the underlying microscopic physical model.
This motivated us to consider the event-by-event fluctuation given in Eq.~\eqref{v2sigma}.
The corresponding results are shown in Tab.~\ref{tabv2varb}.
Compared with the results presented in the foregoing table, the average $v_2$ obtained using MLE is closer to the true value.
This is understood as the particle correlation method estimates $v_n^2$ and contains a fraction of flow fluctuations~\cite{Ollitrault:2009ie, CMS:2013jlh}.
Regarding the variance, we note that the value shown in the last column becomes less than $\sigma^2=0.01$ since, in the Monte Carlos process, all negative harmonic coefficients are discarded.
By definition, the MLE cannot entirely remove the flow fluctuations due to statistical uncertainty.
Nonetheless, in principle, it attains a more accurate value than other estimators, particularly when the multiplicity becomes more significant.

\begin{table}
\caption{The average and variance of the estimated elliptic flow $v_2$ using the two-particle correlation and MLE methods.
The results are obtained using $N=1000$ events and for different multiplicities per event.
The averages $\mu[\cdots]$ and variance Var$[\cdots]$ of the estimators are evaluated on an event-by-event basis.}\label{tabv2vara}
\begin{tabular}{c ccc ccc}
        \hline\hline
        $M $~~&~~$\sqrt{\mu\left[\widehat{v_2^2}\right]}$~~&~~$\mu\left[\hat{v_2}\right]$~~&~~$\quad $~~&~~$\mathrm{Var}\left[\sqrt{\widehat{v_2^2}}\right]$~~&~~$\mathrm{Var}\left[\hat{v_2}\right]$~~&~~$\quad $~~\\ 
        \hline 
        &~~particle correlation~~&~~MLE~~&~~true value~~&~~particle correlation~~&~~MLE~~&~~true value~~\\
        \hline
          500               & 0.1978 & 0.2015 & 0.2 & $9.58 \times 10^{-4} $      & $7.77 \times 10^{-4} $         &   0 \\
        \hline
         1000              & 0.1994 & 0.2021 & 0.2 &  $4.78\times10^{-4}$       &  $3.05 \times 10^{-4}$          &    0 \\
        \hline
         2000             & 0.1997 & 0.2014 & 0.2 &  $2.11 \times10^{-4}$       &  $1.78 \times 10^{-4}$          &     0\\
        \hline\hline
 \end{tabular}
 \end{table}

\begin{table}
\caption{The same as Tab.~\ref{tabv2vara}, but the elliptic flow is generated by including additional Gaussian fluctuations governed by Eq.~\eqref{v2sigma}.}\label{tabv2varb}
\begin{tabular}{c ccc ccc}
        \hline\hline
        $M $~~&~~$\sqrt{\mu\left[\widehat{v_2^2}\right]}$~~&~~$\mu\left[\hat{v_2}\right]$~~&~~$\quad $~~&~~$\mathrm{Var}\left[\sqrt{\widehat{v_2^2}}\right]$~~&~~$\mathrm{Var}\left[\hat{v_2}\right]$~~&~~$\quad $~~\\ 
        \hline 
        &~~particle correlation~~&~~MLE~~&~~true value~~&~~particle correlation~~&~~MLE~~&~~true value~~\\
        \hline 
          500               & 0.2221 & 0.2113 & 0.2 & $9.69 \times 10^{-3} $      & $1.11 \times 10^{-2}$~~&~~$8.90 \times 10^{-3}$ \\
        \hline
         1000              & 0.2252 & 0.2132 & 0.2 &  $8.86 \times10^{-3}$       &  $1.03 \times 10^{-2}$~~&~~$8.90 \times 10^{-3}$ \\
        \hline
         2000             & 0.2252 & 0.2135 & 0.2 &  $8.79 \times10^{-3}$       &  $1.03 \times 10^{-2}$~~&~~$8.90 \times 10^{-3}$\\
        \hline\hline
\end{tabular}
\end{table}

In order to make the above discussions about flow fluctuations more concrete, in Fig.~\ref{figv2v3scatter}, we show the scatter plots for extracted $v_2^2$ and $v_3^2$ for individual events using both methods.
In the calculations, the events are generated whose multiplicities are drawn randomly according to a uniform distribution up to $M=5000$. 
The estimated flow harmonics are shown as a function of event multiplicity.
From Fig.~\ref{figv2v3scatter}, it is observed that the estimations of MLE primarily possess a smaller statistical uncertainty when compared to other methods, as the hollow red triangles are confined mainly within the region determined by the filled black circles. 
These results are consistent with those shown above in Tabs.~\ref{tabv2vara}.
It is also noted that $v_n^2$ derived from the particle correlation might be negative, but the values of $v_n$ estimated by MLE are always found positive in our numerical calculations.

Now, we present the results on the event planes extracted by the MLE method.
Since the particle correlation method does not provide an estimation of the event planes, the obtained results are compared with a conventional event-plane approach~\cite{Poskanzer:1998yz}, adopting the following formula
\begin{eqnarray}
\widehat{\Psi}_n= \frac1n\mathrm{atan}\left(\sum_{j=1}^M\sin n\phi_j, \sum_{j=1}^M \cos n\phi_j\right) .
\label{eqpsin}
\end{eqnarray}
The results are presented in Tab.~\ref{tabPsin_com}.

\begin{table}
\caption{The event planes $\Psi_n$ obtained by MLE and event-plane methods.
The results are obtained for a single event but with different multiplicities.}\label{tabPsin_com}
\begin{tabular}{c ccc ccc ccc}
        \hline\hline
        &\multicolumn{3}{c}{$\Psi_2$}&\multicolumn{3}{c}{$\Psi_3$}&\multicolumn{3}{c}{$\Psi_4$} \\
        \hline
        $M$ &~~500~~&~~1000~~&~~2000~~&~~500~~&~~1000~~&~~2000~~&~~500~~&~~1000~~&~~2000~~ \\
        \hline 
         true value                &~~1.185~~&~~2.943~~&~~2.435~~&~~1.519~~&~~1.695~~&~~1.389~~&~~0.682~~&~~1.272~~&~~0.802~~ \\
        \hline
         event plane method        &~~1.130~~&~~2.986~~&~~2.462~~&~~1.774~~&~~1.844~~&~~1.389~~&~~0.685~~&~~1.242~~&~~0.797~~\\
        \hline
         MLE                       &~~1.134~~&~~2.983~~&~~2.458~~&~~1.849~~&~~1.864~~&~~1.408~~&~~0.659~~&~~1.245~~&~~0.790~~\\
        \hline\hline
 \end{tabular}
 \end{table}

To verify the statistical robustness of the method, we explore the multiplicity and event number dependence of the MLE method.
The results are presented in Tabs.~\ref{tabmultiinc} and~\ref{tabeventinc1000}.
To study the asymptotical normality of the MLE approach, one shows in Tab.~\ref{tabmultiinc} the average and variance of the estimator for the flow harmonics evaluated at different values of multiplicity.
It is shown that as the multiplicity increases, all the estimations of flow harmonics become more precise while the variance decreases, as expected.
On the other hand, Tab.~\ref{tabmultiinc} shows the dependence of the event average and variance of MLE on the event number.
It is indicated that both the average and variance of the estimator become stable as long as the event number is big enough.
This is readily understood as, by definition, the events are independently sampled from the likelihood distribution, whose behavior is therefore governed by the central limit theorem.

\begin{table}
\caption{The dependence of the MLE method on the event multiplicity $M$.
The calculations are carried out with $N=1000$ events.}\label{tabmultiinc}
\begin{tabular}{c c  c  c  }
        \hline\hline
        $M$           &~~500~~&~~1000~~&~~2000~~ \\
        \hline 
          $\mu\left[\hat{v_2}\right]$                  &~~0.205~~&~~0.202~~&~~0.202~~\\
        \hline
         $\mathrm{Var}\left[\hat{v_2}\right]$             & ~~$8.95\times 10^{-4}$~~&~~$4.16\times 10^{-4}$~~&~~$2.27\times 10^{-4}$~~\\
        \hline
          $\mu\left[\hat{v_3}\right]$                  &~~0.0847~~&~~0.0843~~&~~0.0830~~\\
        \hline
         $\mathrm{Var}\left[\hat{v_3}\right]$              &~~$1.11\times 10^{-3}$~~&~~$4.32\times 10^{-4}$~~&~~$1.95\times 10^{-4}$~~\\
        \hline
          $\mu\left[\hat{v_4}\right]$                 &~~0.258~~& 0.252~~&~~0.252~~\\
        \hline
         $\mathrm{Var}\left[\hat{v_4}\right]$             &~~$8.99\times 10^{-4}$~~&~~$4.31\times 10^{-4}$~~&~~$2.15\times 10^{-4}$~~\\
        \hline\hline
\end{tabular}
\end{table}
 
\begin{table}
\caption{The dependence of the MLE method on the number of events $N$.
The calculations are carried out with multiplicity $M=1000$.}\label{tabeventinc1000}
 \begin{tabular}{c c  c  c  }
        \hline\hline
        $N$          &~~500~~&~~1000~~&~~2000~~ \\
        \hline
          $\mu\left[\hat{v_2}\right]$                  &~~0.2023~~&~~0.2022~~&~~0.2026~~\\
        \hline
         $\mathrm{Var}\left[\hat{v_2}\right]$            &~~$4.08\times 10^{-4}$~~&~~$4.27\times 10^{-4}$~~&~~$4.41\times 10^{-4}$~~\\
        \hline\hline
 \end{tabular}
 \end{table}

Before closing this subsection, we note that since the MLE evaluates the flow harmonics in a way that is independent of specific particle combinations, it naturally furnishes certain harmonic products that are not obtained straightforwardly from the particle correlation analysis.
Moreover, owing to the invariance property of MLE, a product of the MLEs is the MLE of the product of harmonic coefficients.
We will further address the last point in the following subsection in the context of detector inefficiency.

\subsection{MLE and detector inefficiency}\label{section4}

In practice, a detector's acceptance might not be uniform, potentially leading to non-negligible systematic bias in anisotropic flow analysis.
In terms of particle correlation and Q-vectors, this issue has been addressed by various authors~\cite{Bilandzic:2013kga}.
This subsection explores how the MLE method can be applied to scenarios involving detector inefficiency. 
In particular, we elaborate on a scheme to compensate for the detector's nonuniform acceptance by incorporating {\it weight} into the formalism of the likelihood function.
The proposed scheme is then implemented and illustrated by considering two specific detector inefficiencies.
Moreover, we show that MLE can be employed to evaluate certain harmonic products that are not straightforwardly embraced by the particle correlation approach.

One may quantify the nonuniformity by the detector's acceptance rate as a function of the azimuthal angle, $g(\phi)\le 1$, where $g(\phi)=1$ corresponds to a perfect detector.
For simplicity, we also ignore the dependence of the acceptance on transverse momentum and rapidity.
For the MLE scheme, the likelihood function can be modified to adapt to this inefficiency by including a weight factor defined by $w(\phi)=1/g(\phi)$.
Subsequently, one generalizes Eqs.~\eqref{eqlikelihood} and~\eqref{eqlogl} to the form
\begin{eqnarray}
{\ell}'(\theta;\phi_{1},\phi_{2},\phi_{3} \cdots \phi_{{\rm M}}  )=\log\mathcal{L}'(\theta;\phi_{1},\phi_{2},\phi_{3} \cdots \phi_{{\rm M}}  )
=\sum_{j=1}^{M}w(\phi_j)\log f_1(\phi_j;\theta).
\label{eqlikewk}
\end{eqnarray}
It is noted that the weight factor gives a correction in the power of the one-particle distribution function, compensating for the suppressed multiplicity.
Using Eq.~\eqref{eqlikewk}, the MLE calculations remain essentially unchanged.
 
As mentioned above, the particle correlation method is mostly applicable to a specific class of set of correlator Eq.~\eqref{nkCorr} where $\sum_{j=1}^k n_j=0$.
In other words, the quantity is isotropic because the event planes cancel out entirely in the remaining expression. 

In what follows, we illustrate the scheme through two specific forms of detector acceptance.
By employing the MLE method, we consider the following products of flow harmonics
\begin{eqnarray}
 \langle 2 \rangle&\equiv&\langle 2\rangle_{-2, 2}=v_{2}^2=0.04 ,\nonumber\\
 \langle 3 \rangle&\equiv&\langle 3\rangle_{-2, -2, 4}=v_{2}^2v_{4}=0.01 , \nonumber\\
 \langle 4 \rangle&\equiv&\langle 4\rangle_{-3, -2, 2, 3}=v_{2}^2v_{3}^2=0.000256 ,\nonumber\\
 \langle 2' \rangle&\equiv&v_2v_3=0.016,\nonumber\\
 \langle 3' \rangle&\equiv&v_2^2v_3=0.0032,\nonumber\\
 \langle 4' \rangle&\equiv&v_2^2v_3v_4=0.0008 ,
\label{eqvnvalueaso}
\end{eqnarray}
where the last equality gives the true values.
It is noted that the first three quantities are essentially multi-particle correlators of the form Eq.~\eqref{nkCorr}, which can also be evaluated~\footnote{As in~\cite{Bilandzic:2013kga}, the second row in Eqs.~\eqref{eqvnvalueaso} is obtained by considering $\Psi_2=\Psi_4$.} using the particle correlation scheme as in~\cite{Bilandzic:2013kga}.
For instance, the event average $\langle 2\rangle_{-2, 2}$ vanishes due to the event average carried out for the terms involving the event planes. 
However, the last three quantities in Eq.~\eqref{eqvnvalueaso} do not fit into this category as the corresponding correlator vanishes.
Specifically, using the MLE approach, one estimates $v_n$ and their products for individual events, and the event averages $\langle v_2^2v_4 \rangle$ and $\langle v_2^2v_3 \rangle$ can be evaluated in a similar fashion irrelevant to any specific particle tuple. 

Our first example is a simplified scenario according to~\cite{Bilandzic:2010jr}, where the detector's acceptance is given by a piecewise function
\begin{eqnarray}
g_1(\phi)= \begin{cases}
0.5\ \quad &\pi/3 <\phi\leq 2\pi/3 \\
1.0\ \quad &\mathrm{ otherwise}
\end{cases} .
\label{eqeffpiecef}
\end{eqnarray}

\begin{figure}[ht]
\begin{tabular}{cc}
\vspace{-26pt}
\begin{minipage}{250pt}
\centerline{\includegraphics[width=1.2\textwidth,height=1.0\textwidth]{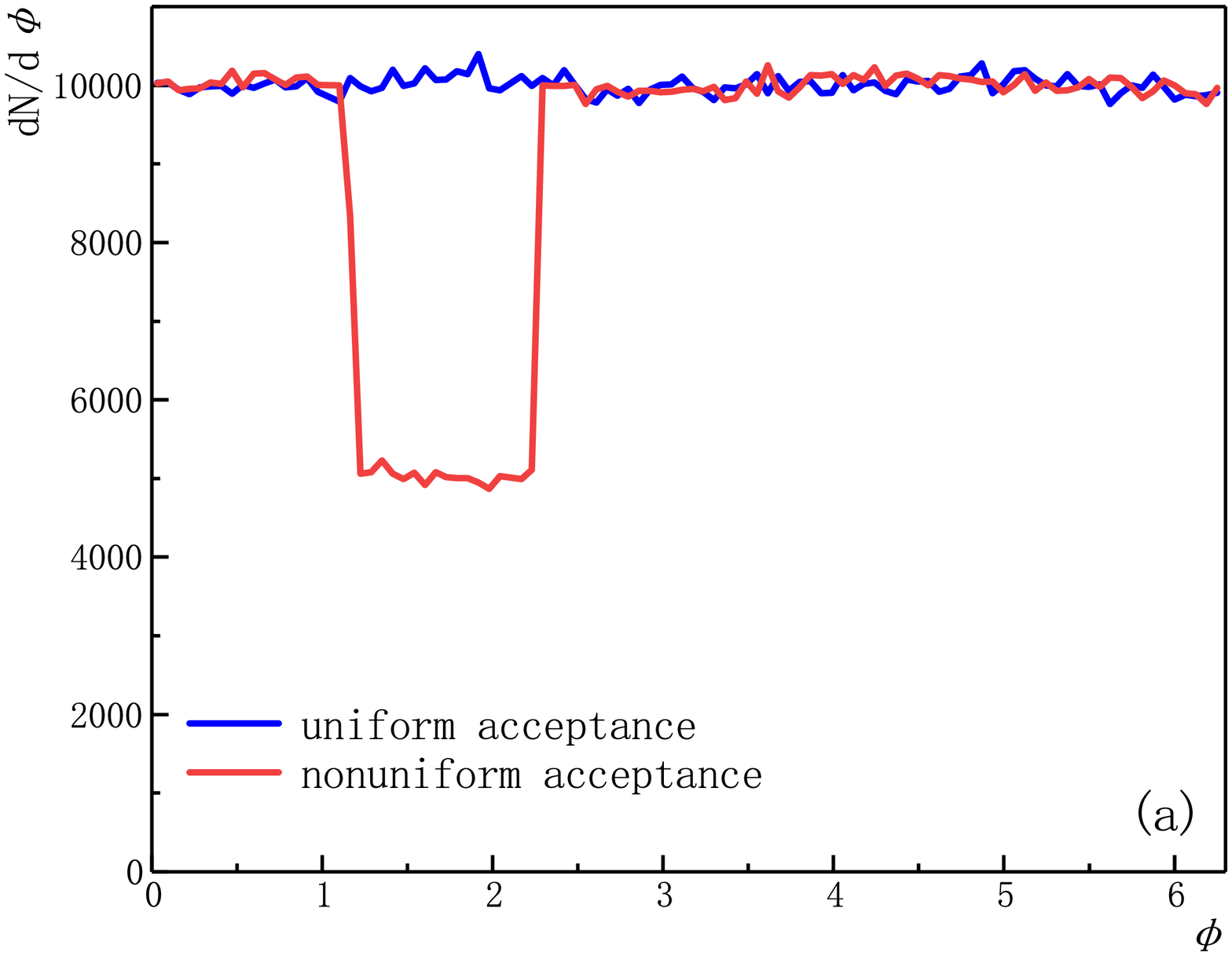}}
\end{minipage}
&
\begin{minipage}{250pt}
\centerline{\includegraphics[width=1.2\textwidth,height=1.0\textwidth]{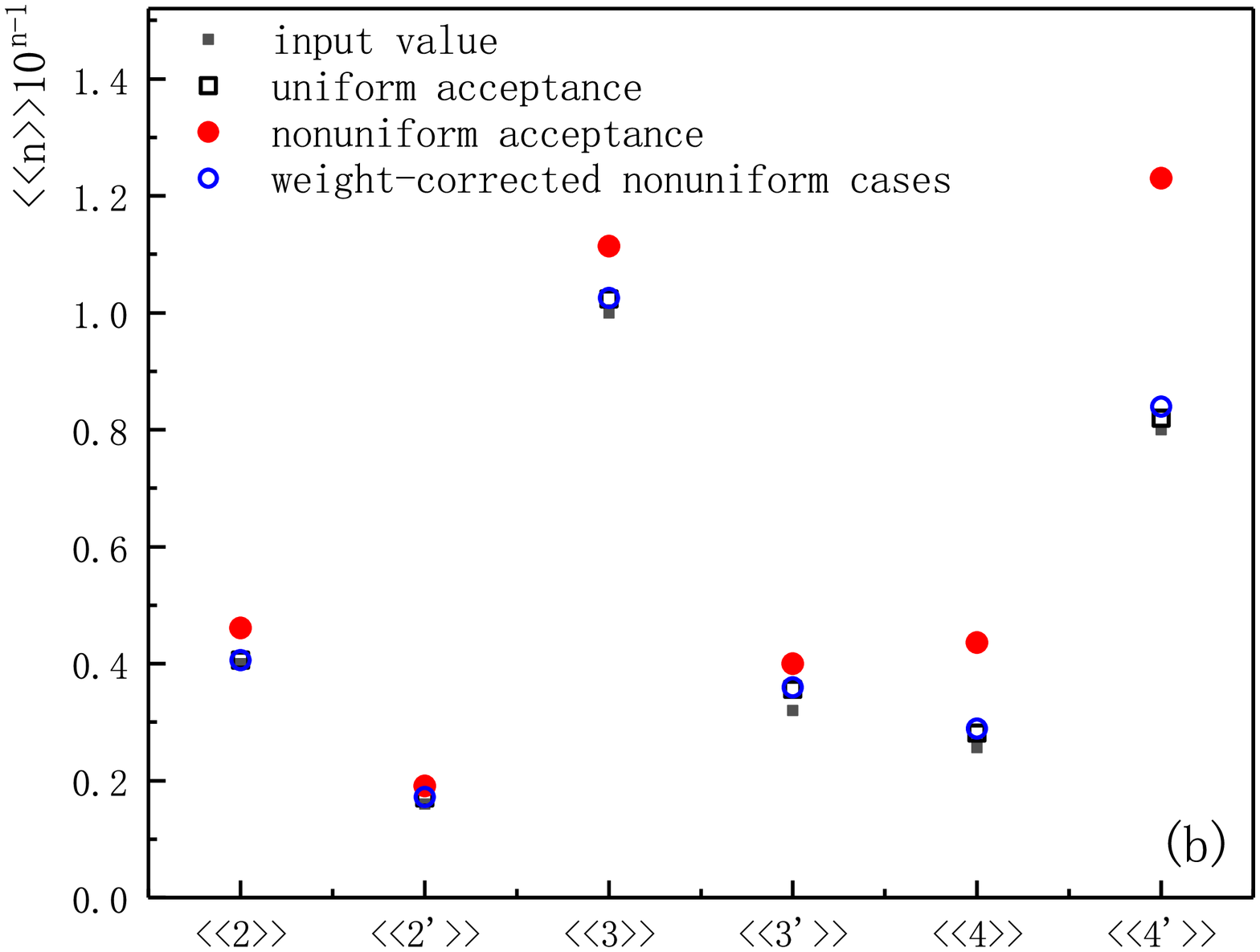}}
\end{minipage}
\\
\vspace{22pt}
\end{tabular}
\vspace{12pt}
\renewcommand{\figurename}{Fig.}
\caption{(Color Online) The results for a simplified detector's acceptance given by a step function Eq.~\eqref{eqeffpiecef}.
(a): The particle generated and observed by the detector with a uniform (blue line) and nonuniform (red line) azimuthal acceptance.
(b): The resultant products of flow harmonics for uniform, nonuniform, and weight-corrected nonuniform cases.}
\label{piecefcorr}
\end{figure}

\begin{figure}[ht]
\begin{tabular}{cc}
\vspace{-26pt}
\begin{minipage}{250pt}
\centerline{\includegraphics[width=1.2\textwidth,height=1.0\textwidth]{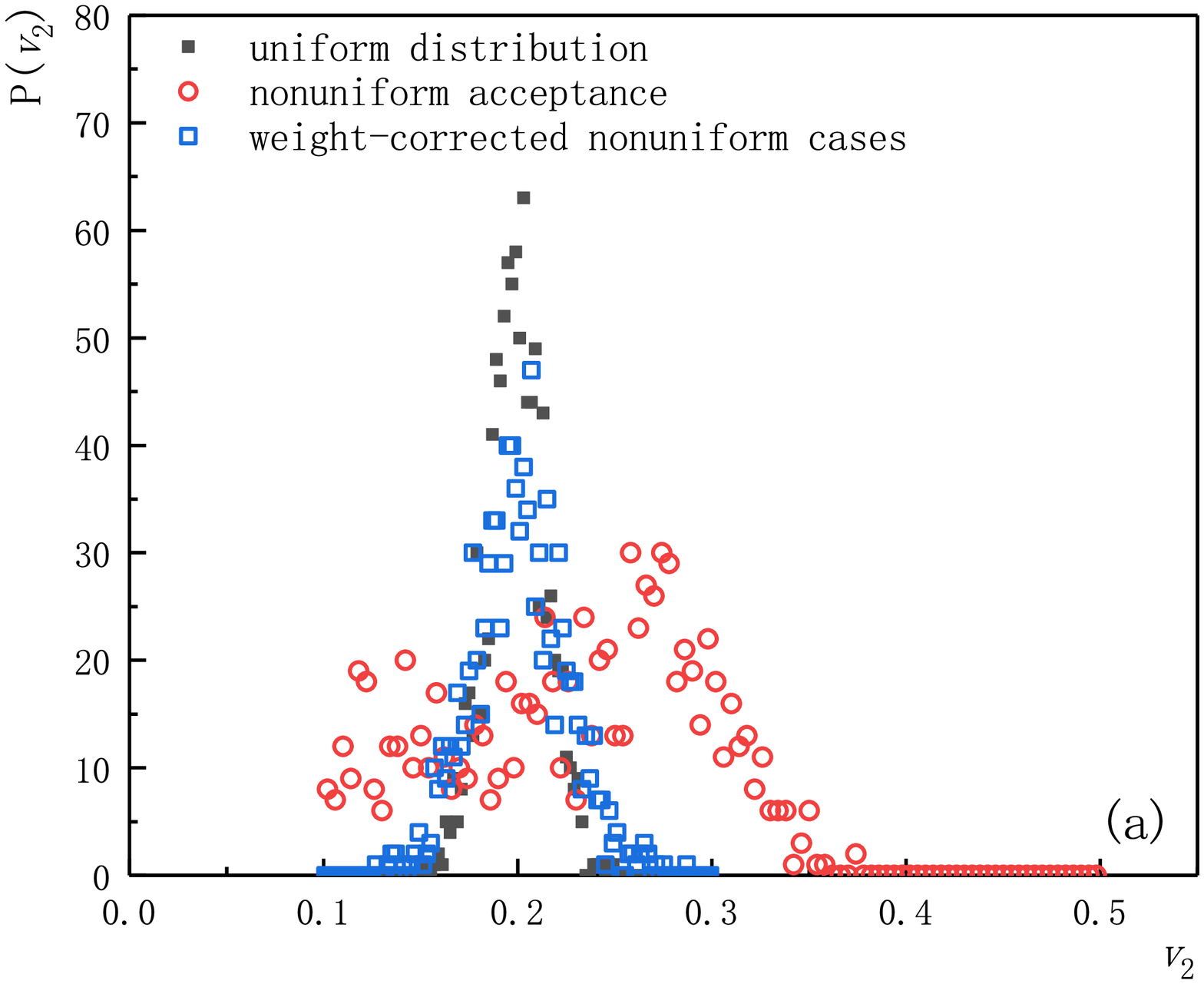}}
\end{minipage}
&
\begin{minipage}{250pt}
\centerline{\includegraphics[width=1.2\textwidth,height=1.0\textwidth]{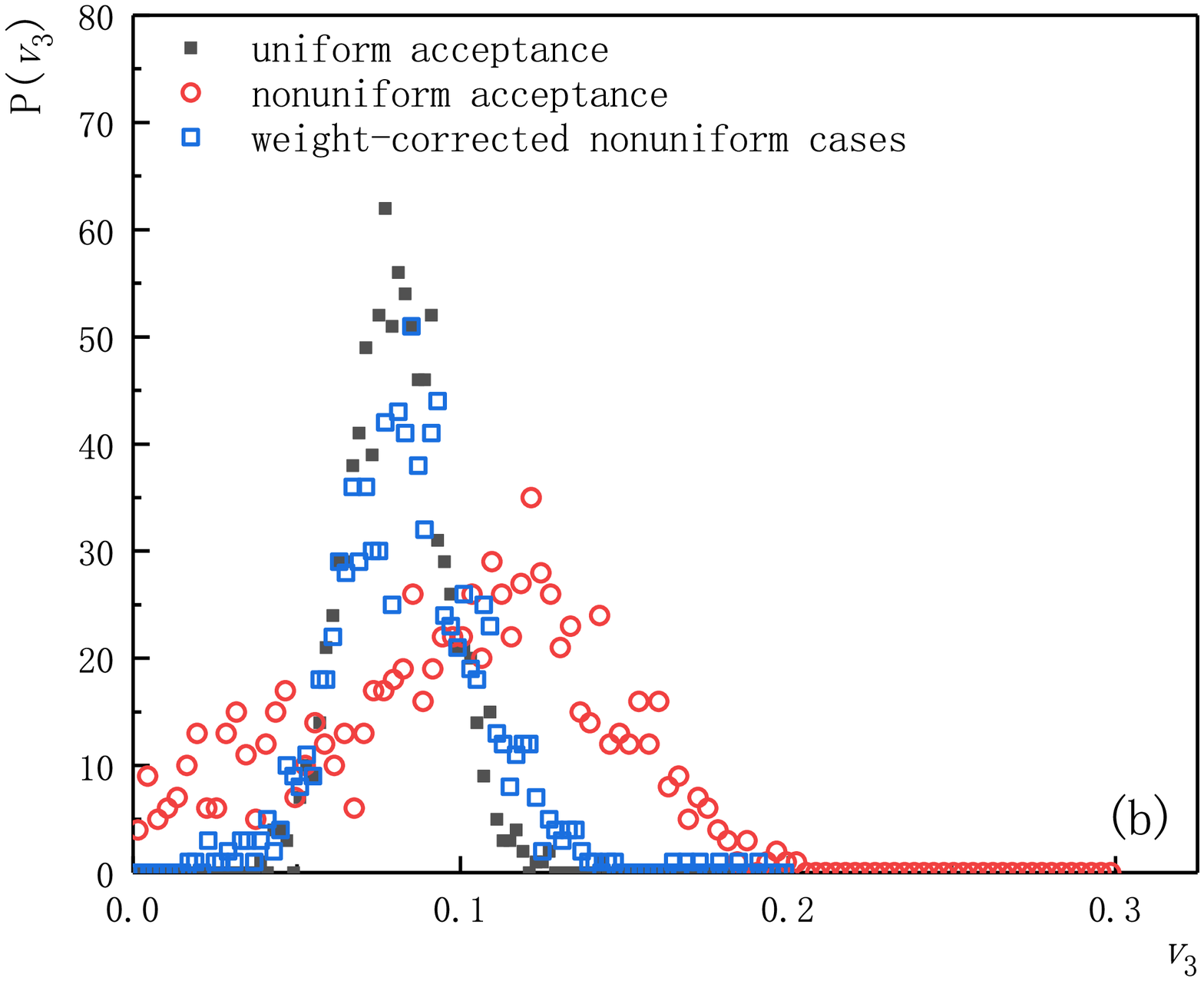}}
\end{minipage}
\end{tabular}
\vspace{12pt}
\renewcommand{\figurename}{Fig.}
\caption{(Color Online) The distributions of the estimated $v_2$ and $v_3$ on an event-by-event basis using the MLE method, evaluated for the detector's acceptance given by Eq.~\eqref{eqeffpiecef}.}
\label{eff1corr234}
\end{figure}

The results are shown in Figs.~\ref{piecefcorr} and~\ref{eff1corr234}.
In the left plot of Fig.~\ref{piecefcorr}, we present the distributions of the detected particles according to the Monte Carlo procedure.
These particles are then utilized by the MLE method to estimate the quantities given in Eq.~\eqref{eqvnvalueaso}.
As shown on the right plot of Fig.~\ref{piecefcorr}, for all six quantities, the MLE with weight is shown to give appropriate compensation for the detector inefficiency.
The results of the first three quantities defined in Eq.~\eqref{eqvnvalueaso} are mainly consistent with the results obtained by using Q-vectors~\cite{Bilandzic:2013kga}.
For the last three quantities, where reasonable corrections are also achieved, the evaluation using MLE is somewhat unique in its own right.
We also show in Fig.~\ref{eff1corr234} the distributions of the estimated $v_2$ and $v_3$ on an event-by-event basis.
For both cases, it is demonstrated clearly that the expected value and probability distributions of the relevant quantities are properly recuperated.

\begin{figure}[ht]
\begin{tabular}{cc}
\vspace{-26pt}
\begin{minipage}{250pt}
\centerline{\includegraphics[width=1.2\textwidth,height=1.0\textwidth]{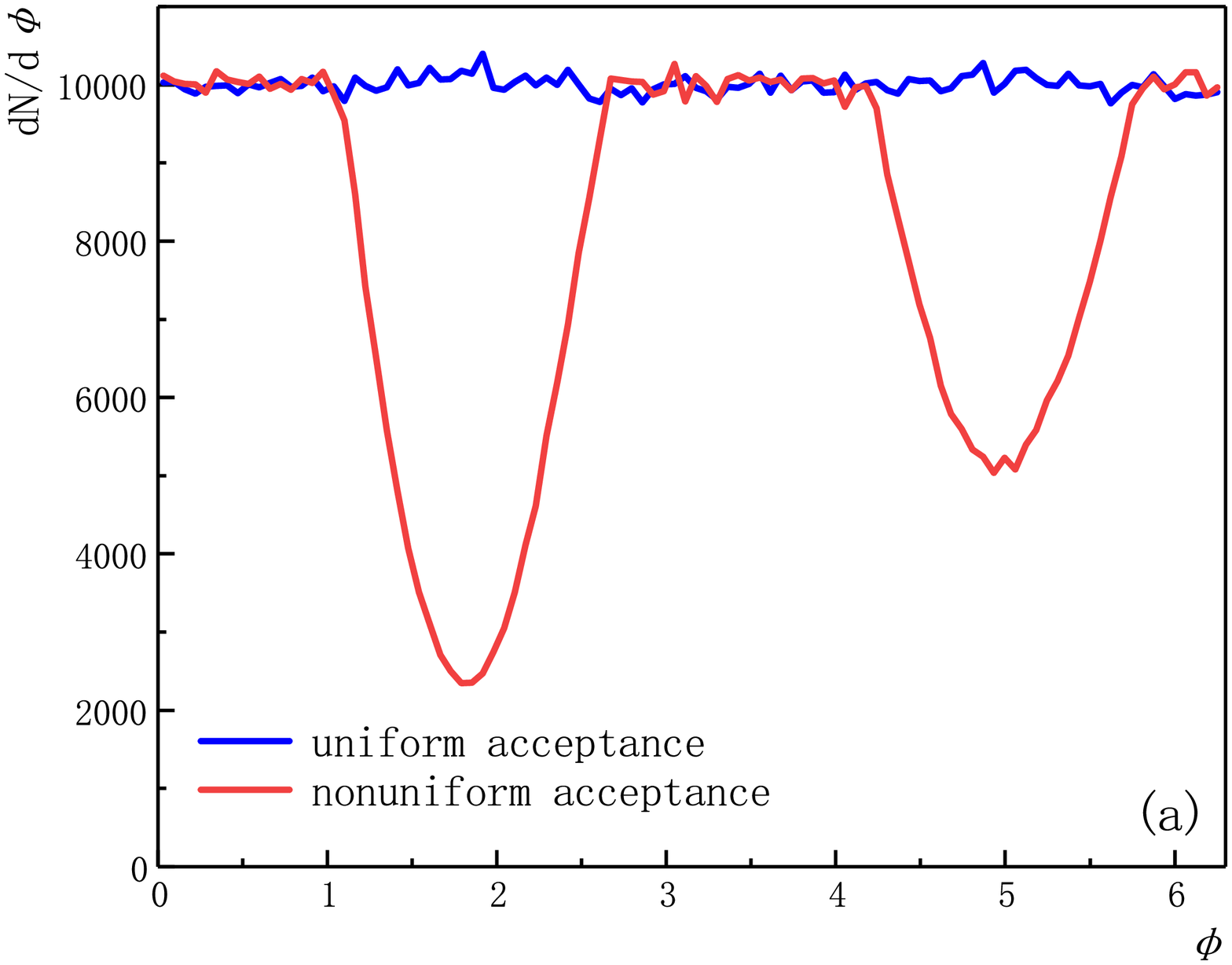}}
\end{minipage}
&
\begin{minipage}{250pt}
\centerline{\includegraphics[width=1.2\textwidth,height=1.0\textwidth]{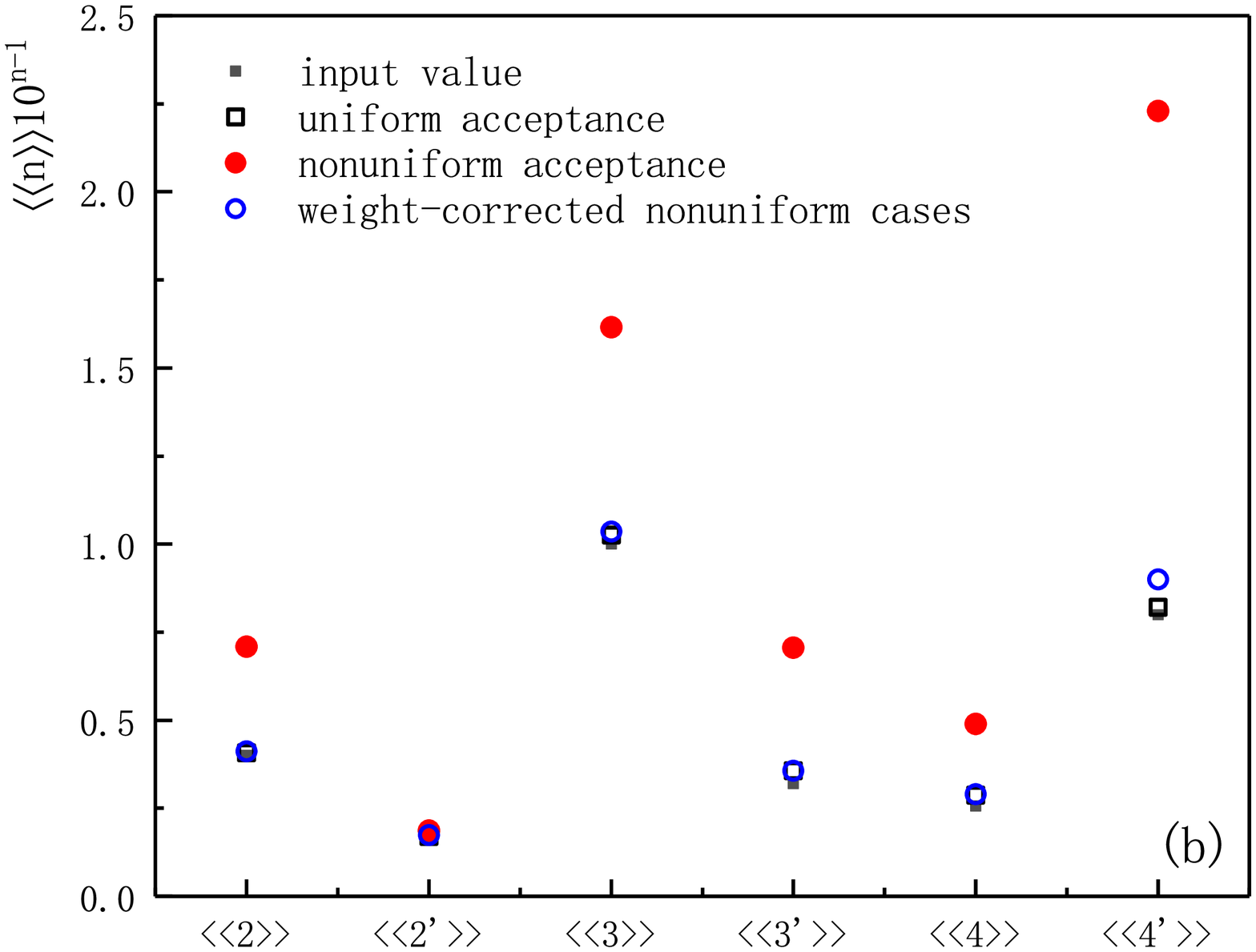}}
\end{minipage}
\end{tabular}
\vspace{12pt}
\renewcommand{\figurename}{Fig.}
\caption{(Color Online) The same as Fig.~\ref{piecefcorr}, but for a more realistic nonuniform acceptance given by Eq.~\eqref{eqeff2}.
(a): The particle generated and observed by the detector with a uniform (blue line) and nonuniform (red line) azimuthal acceptance.
(b): The resultant products of flow harmonics for uniform, nonuniform, and weight-corrected nonuniform cases.}
\label{figeff2}
\end{figure}

\begin{figure}[ht]
\begin{tabular}{cc}
\vspace{-26pt}
\begin{minipage}{250pt}
\centerline{\includegraphics[width=1.2\textwidth,height=1.0\textwidth]{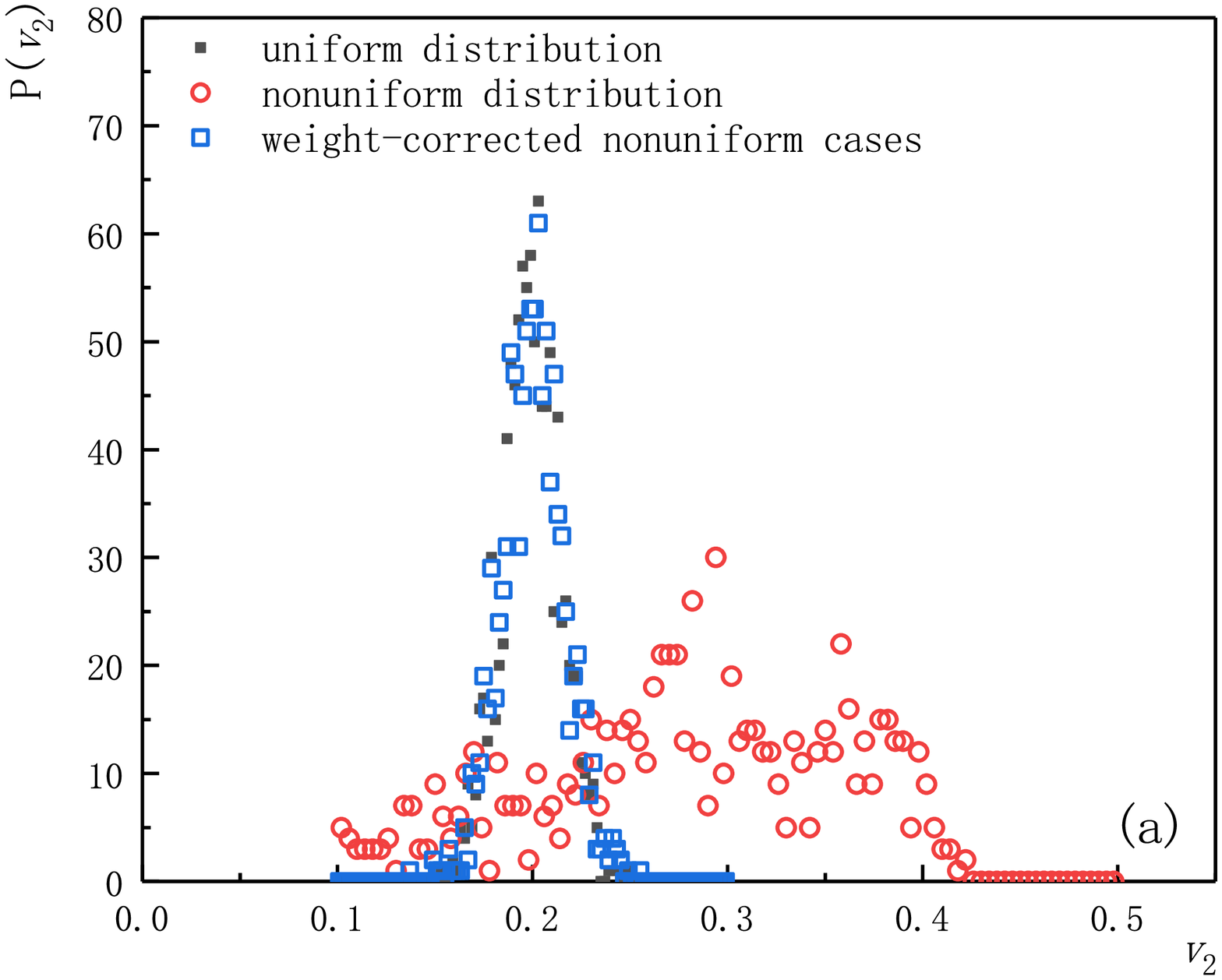}}
\end{minipage}
&
\begin{minipage}{250pt}
\centerline{\includegraphics[width=1.2\textwidth,height=1.0\textwidth]{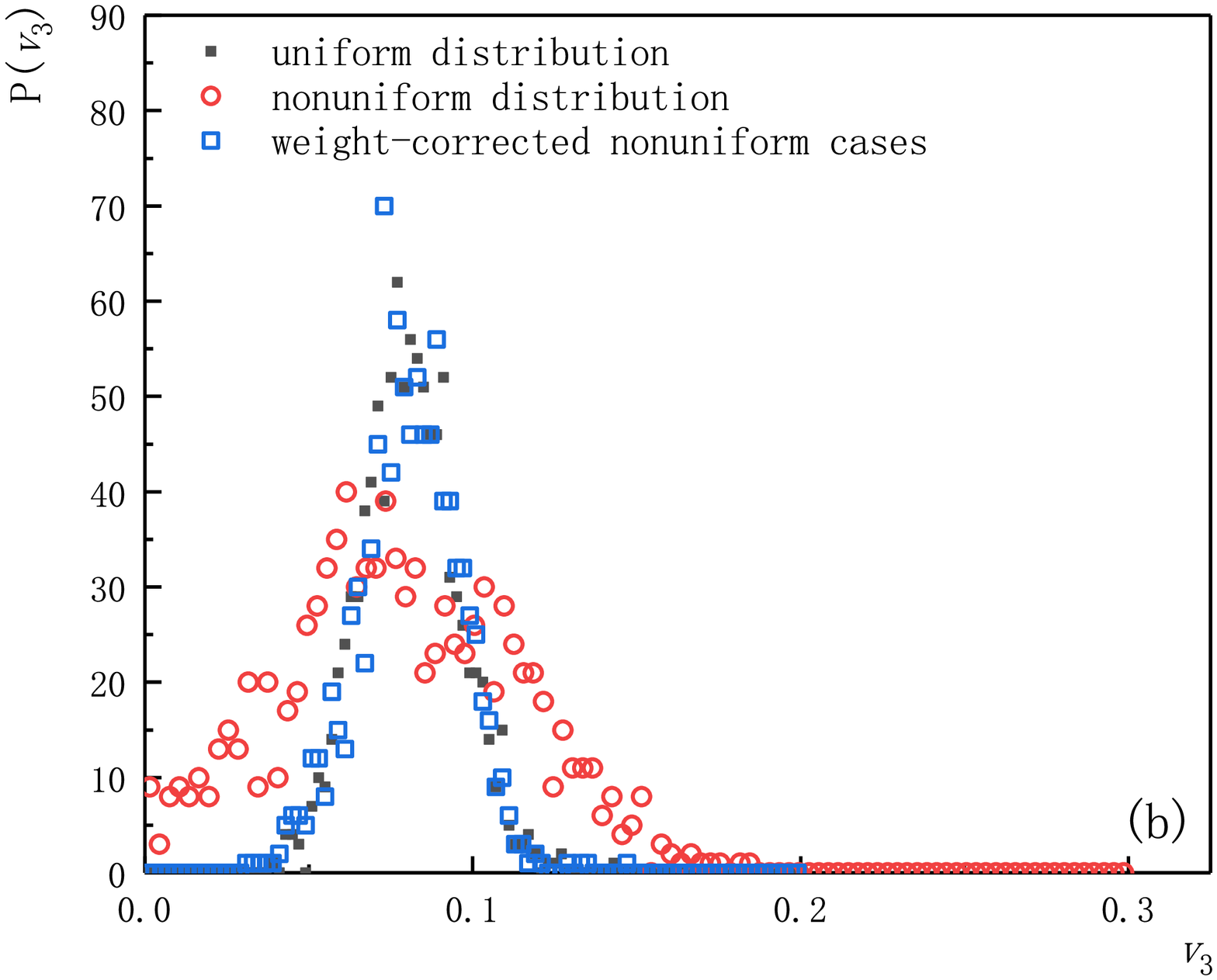}}
\end{minipage}
\end{tabular}
\vspace{12pt}
\renewcommand{\figurename}{Fig.}
\caption{(Color Online) The same as Fig.~\ref{eff1corr234}, but for the detector's acceptance given by Eq.~\eqref{eqeff2}.
}
\label{figv234corr}
\end{figure}

We proceed to elaborate on a more realistic detector's acceptance function given by the following form
\begin{eqnarray}
g_2(\phi)= \begin{cases}
1+\mathrm{exp}(-\phi/7)\sin{2(\phi+0.5)}\ \quad &1.07 <\phi\leq 2.64~\mathrm{or}~4.21 <\phi\leq 5.78\\
1.0\ \quad &\mathrm{ otherwise}
\end{cases}  .
\label{eqeff2}
\end{eqnarray}
The above acceptance's function mimics a realistic detector where the acceptance is suppressed at the forward and backward directions~\cite{NA49:2003njx}.  
Fig.~\ref{figeff2} shows the distribution of the detected particles and the resulting estimation of harmonic products.
Also, Fig.~\ref{figv234corr} gives the distribution of the estimations for $v_2$ and $v_3$ regarding individual events.
Again, it is clearly seen that the MLE method does a reasonable job of estimating these quantities and their probability distributions while coping with detector inefficiency, particularly for the ones which cannot be furnished straightforwardly by the particle correlation method.

\subsection{A fictitious scenario with fluctuating event plane}\label{section5}

In this section, we elaborate on a fictitious scenario where the event plane in Eq.~\eqref{oneParDis} is not a well-defined quantity in the distribution function.
In particular, we consider that the event plane is subjected to a normal distribution
\begin{eqnarray}
\Psi_2 \sim N(\mu, \sigma^2) .
\label{psi2sigma}
\end{eqnarray}
where $\mu$ is sampled uniformly from the interval $\mu\in (0, \pi)$ and $\sigma^2=\pi/12$ or $\pi/24$.

Using the two-particle correlation estimator Eq.~\eqref{Estimatorv2c2} and MLE, the results are presented in Tab.~\ref{tabpsi2varc}.
Since a parameter of the distribution function, i.e., the event plane, is governed by a probability distribution, the estimation scheme should be taken with a grain of salt.
As expected, the estimated $v_2$ becomes less accurate as the deviation $\sigma^2$ becomes more significant.
Nonetheless, the estimations of $v_2$ are numerically convergent for both approaches and are mostly reasonable when the fluctuations are not significant.
When compared with two-particle correlation, MLE yields essentially better results.
Specifically, if one assumes that $\Psi_2$ is a fixed parameter in the one-particle distribution, MLE gives slightly better results than those from particle correlation.
However, if one considers that $\Psi_2$ is generated by a distribution where the variance $\sigma^2$ is taken as an unknown parameter as the remaining ones, MLE (dubbed as ``MLE-mod'' in Tab.~\ref{tabpsi2varc}) yields better results.
We note that such an extension does not seem straightforward in the case of the particle-correlation estimator.

\begin{table}
\caption{The estimated elliptic flow using the particle correlation and MLE estimators where the event-plane $\Psi_2$ is generated by considering additional Gaussian fluctuations given by Eq.~\eqref{psi2sigma}.}\label{tabpsi2varc}
\begin{tabular}{c cccc | cccc}
        \hline\hline
        $\sigma^2$ & \multicolumn{4}{c}{$\pi/12$} & \multicolumn{4}{c}{$\pi/24$} \\
        \hline 
        $M $ &~~particle correlation~~&~~MLE~~&~~MLE-mod~~&~~true value~~&~~particle correlation~~&~~MLE~~&~~MLE-mod~~&~~true value~~\\
%        ~~&~~$\sqrt{\mu\left[\widehat{v_2^2}\right]}$~~&~~$\mu\left[\hat{v_2}\right]$~~&~~$\quad $~~&~~$\mathrm{Var}\left[\sqrt{\widehat{v_2^2}}\right]$~~&~~$\mathrm{Var}\left[\hat{v_2}\right]$~~&~~$\quad $~~\\ 
%        \hline
%          50        & 0.1166   & 0.1086   & 0.1095    & 0.2 &  0.1459      & 0.1496   & 0.1456       &    0.2 \\
        \hline
         100        & 0.1174   & 0.1227   & 0.1805    & 0.2 &  0.1501      & 0.1554   & 0.2010       &    0.2 \\
        \hline
         500        & 0.1171   & 0.1198   & 0.1794    & 0.2 &  0.1556      & 0.1573   & 0.2156       &    0.2\\
        \hline\hline
 \end{tabular}
 \end{table}

\section{Concluding remarks}\label{section6}

In this work, we study the possibility of employing the MLE as a flow estimator.
The proposed estimator possesses three interesting features.
First, due to its asymptotical normality, MLE provides a smaller variance when comparing other estimators.
Since a part of flow fluctuations comes from the finite multiplicity of realistic collision events, this feature can be utilized to suppress undesirable statistical uncertainties.
Second, the proposed approach provides a means to access specific mixed harmonics, which cannot be straightforwardly obtained using methods primarily based on particle correlations. 
Third, the MLE flow estimator is robust for the fictitious scenario where some parameters of the probability distribution are not well-defined. 
The resultant flow harmonics obtained using MLE were compared with those derived using other existing methods.
The obtained results were analyzed using the Wald, likelihood ratio, and score tests of hypotheses.
We also explore the dependence of extracted flow harmonics on the event multiplicity and number of events.
Moreover, it is shown that the proposed approach works efficiently to deal with the deficiency in detector acceptability.
Therefore, it is argued that the MLE furnishes a meaningful alternative to the existing approaches.

We also acknowledge that the proposed scheme encounters certain limitations or drawbacks.
First, it is computationally expensive in its present form.
The algorithm complexity of the MLE approach increases essentially linearly with the multiplicity, similar to that of the particle-correlation method.
Besides, the MLE's computational time increases geometrically with the dimension of the parameter space.
Second, although MLE is asymptotically normal, unbiased, and consistent, which generally attains more accurate values than other estimators, numerical results also seem to reveal a certain degree of bias at finite multiplicity.
To our knowledge, mathematically confirming whether the flow MLE is a biased estimator is not straightforward and has not been addressed in this study.
Lastly, the present study has been primarily focused on the integrated flow.
To generalize the scheme to differential flow may also face further challenges such as, among others, the computational feasibility.
Besides, it is interesting to examine the performance of MLE when non-flow plays a significant role.
We plan to explore these topics in future studies.

\section*{Acknowledgements}

We are thankful for enlightened discussions with Sandra Padula, Mike Lisa, Matthew Luzum, and Giorgio Torrieri.
We appreciate the anonymous referee's valuable comments and suggestions. 
WLQ expresses gratitude towards the Institute for Nuclear Theory at the University of Washington for their generous hospitality during the execution of this work, documented under report no. INT-PUB-23-017.
This work is supported by the National Natural Science Foundation of China.
We also gratefully acknowledge the financial support from Brazilian agencies 
Funda\c{c}\~ao de Amparo \`a Pesquisa do Estado de S\~ao Paulo (FAPESP), 
Funda\c{c}\~ao de Amparo \`a Pesquisa do Estado do Rio de Janeiro (FAPERJ), 
Conselho Nacional de Desenvolvimento Cient\'{\i}fico e Tecnol\'ogico (CNPq), 
and Coordena\c{c}\~ao de Aperfei\c{c}oamento de Pessoal de N\'ivel Superior (CAPES).
A part of this work was developed under the project Institutos Nacionais de Ciências e Tecnologia - Física Nuclear e Aplicações (INCT/FNA) Proc. No. 464898/2014-5.
This research is also supported by the Center for Scientific Computing (NCC/GridUNESP) of S\~ao Paulo State University (UNESP).
This work is also supported by the Postgraduate Research \& Practice Innovation Program of Jiangsu Province under Grant No. KYCX22-3453.

\bibliographystyle{h-physrev}
\bibliography{references_MLE, references_qian}
%\bibliography

\end{document}